\documentclass[conference]{IEEEtran}
\IEEEoverridecommandlockouts
\usepackage{cite}
\usepackage{amsmath,amssymb,amsfonts}
\usepackage{algorithmic}
\usepackage{graphicx}
\usepackage{textcomp}
\usepackage{xcolor}
\usepackage{dblfloatfix}
\usepackage{cancel}
\def\BibTeX{{\rm B\kern-.05em{\sc i\kern-.025em b}\kern-.08em
    T\kern-.1667em\lower.7ex\hbox{E}\kern-.125emX}}
\begin{document}

\title{Nonlinear Stability Assessment Of Type-4 Wind Turbines During Unbalanced Grid Faults Based On Reduced-Order Model}

\author{
\IEEEauthorblockN{Sujay Ghosh, Mohammad Kazem \\ Bakhshizadeh, and~Łukasz Kocewiak}
\IEEEauthorblockA{\textit{Ørsted Wind Power,
 Nesa Allé 1, 2820, Denmark}\\
Email: sujgh@orsted.com}
\and
\IEEEauthorblockN{Guangya Yang}
\IEEEauthorblockA{\textit{Technical University of Denmark},\\ \textit{Anker Engelunds Vej 1, 2800, Denmark}}
}


\maketitle

\begin{abstract}
As the number of converter-based renewable generations in the power system is increasing, the inertia provided by the synchronous generators is reducing, which in turn is reducing the stability margins of the power system. In order to assess the large-signal stability, it is essential to model the wind power plant connections accurately. However, the actual EMT models are often unavailable, black-boxed, or computationally too heavy to model in detail. Hence, simplified reduced-order models (ROMs) resembling the actual system behaviour have gained prominence in stability studies. In this regard, an improved WT ROM was proposed to investigate large signal stability during unbalanced grid faults. The methodology presents a systematic way to model the coupled sequence components of the WT ROM for various grid faults. Based on the studies carried out in this paper, it is observed that post unbalanced grid disturbances the proposed WT ROM correctly tracks the angle and frequency, and its trajectory is a good match when compared to a detailed simulation model in PSCAD.
\end{abstract}

\begin{IEEEkeywords}
Wind Turbine, Reduced order model, Unbalanced grid faults, large signal stability
\end{IEEEkeywords}

\section{Introduction}
The growth in the renewable energy industry is expected to remain strong considering the technological innovations, economies of scale and policy support worldwide. As the number of converter-based renewable generations in the power system is increasing \cite{b1}, the inertia provided by the synchronous generators is reducing, which in turn is reducing the stability margins of the power system. As a result, the system operators are concerned about their respective systems' stable and reliable operation, and have started to impose strict grid code requirements.

Traditionally, the large-signal stability (better known as transient stability) of wind power plant (WPP) connections has been assessed by running batches of time-domain simulations at an operating point of interest. Wherein a detailed wind turbine (WT) model, the low-voltage WPP network and the interconnected AC grid are required. Research has shown that the transient behaviour of grid-following converters deteriorate when connected to weaker grids, which can potentially lead to unstable phenomena such as power system oscillations \cite{b2}\cite{b3}. Furthermore, in \cite{b5}, it has been shown that under large disturbances, the destabilisation of the phase-locked loop (PLL) controls, primarily influences the large-signal stability of the WT system. 

In order to assess the large-signal stability, it is essential to model the WPP connections accurately. However, the actual EMT models are often unavailable, black-boxed, or computationally too heavy to model in detail. Simulating scenarios and contingency analysis through manufacturer-provided EMT models for an entire WPP takes a long time, and stability can only be observed for the predefined events simulated. Hence, simplified reduced-order models (ROMs) resembling the actual system behaviour have gained prominence in stability studies \cite{b6}-\cite{b10}. Such WT ROMs have been designed considering a detailed PLL control model with an approximated control and power system network. 

In general, the WT ROMs in \cite{b6}-\cite{b9} address some nonlinear characteristics, such as the low-frequency PLL dynamics and the influence of weak grids on the dynamics behaviour. However, these ROMs only exhibit autonomous system behaviour, i.e. they do not consider the post-disturbance jumps in initial conditions, the post-fault active current ramp rate, and ramp change in grid frequency, among others. In this regard, as a part of our ongoing research on nonlinear models and stability assessment, an improved WT ROM was proposed in \cite{b10}, addressing all the enhancements. However, the model in \cite{b10} was developed for balanced disturbances, such as three-phase to-ground faults. In reality, unbalanced faults are the most common disturbances; in certain conditions, they may be more severe than balanced faults. Therefore, ROMs must also replicate unbalanced system behaviour. This paper aims to present a methodology to extend the existing improved WT ROM for unbalanced grid conditions. 

\begin{figure*}[h]
    \centering
    \includegraphics[width=16.0cm]{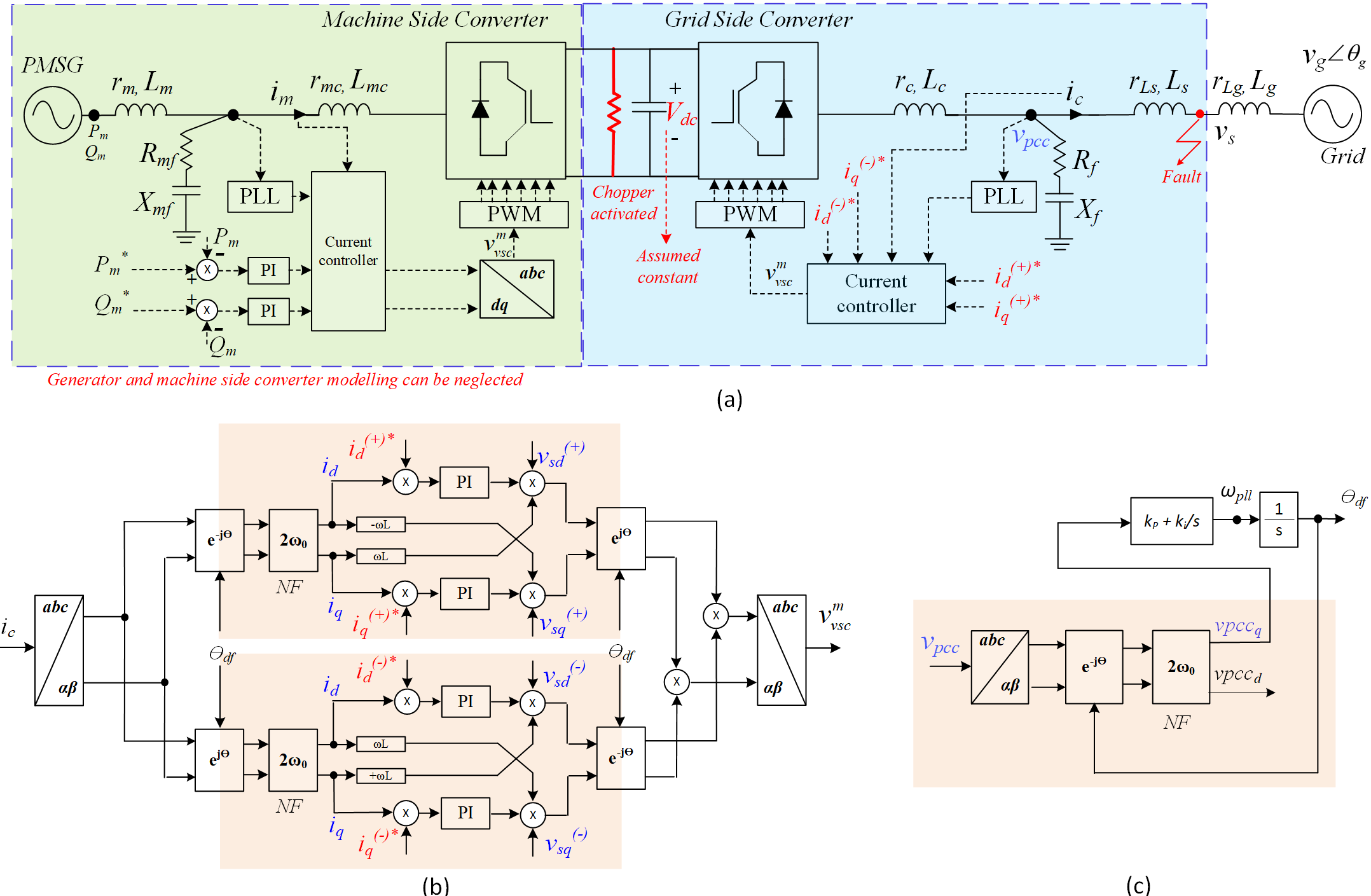}
    \caption{Wind turbine model: (a) Full topology of Type-4 wind turbine system, highlighting the actions/assumptions during faults. (b) Positive and negative sequence current controller for the grid-side WT converter (c) Notch filter-based SRF PLL for synchronisation.}
    \label{FullTop}
\end{figure*}

The PLL is one of the best control systems to synchronise the converter to the grid. Regardless of its ease of implementation, the PLL can be customised to deal with disturbances such as dc offset, harmonics, and unbalanced grid voltage sags \cite{b11}-\cite{b13}. Under unbalanced grid faults, the second harmonic oscillations (2$\omega_0$) can be observed, which cannot be entirely attenuated by PI controllers, and causes steady-state errors. In order to improve the control performance, these oscillations should be cancelled to achieve full control of injected positive and negative sequence currents. This task is carried out by integrating additional filters inside the control loop of a standard PLL structure and current controller \cite{b14}\cite{b15}, such as moving average filter, doubly decoupled filter, and notch filter, among others. In our application a notch filter is considered inside the control loops. 

The paper is organised as follows. Section II presents the modelling of a type-4 wind turbine suitable for large-signal stability analysis under unbalanced grid faults. Section III presents the time domain verification of the proposed method against PSCAD simulations. Finally, Section IV presents the conclusions and future work.

\section{Modelling of Type-IV wind turbine}
This section presents the WT ROM to investigate large signal stability during unbalanced grid faults. Furthermore, it discusses the standard grid code requirements under unbalanced fault conditions.

\subsection{Wind turbine model for balanced grid faults}
\begin{figure}[h]
    \centering
    \includegraphics[width=9.0cm]{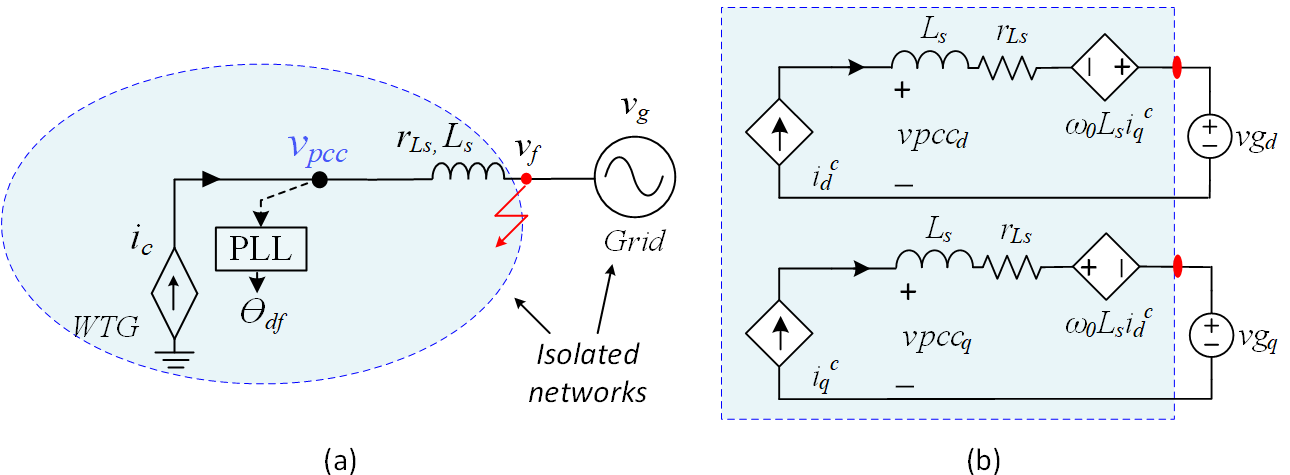}
    \caption{Wind turbine model: (a) Reduced order model (ROM) of the Type-4 wind turbine considering the actions/assumptions. (b) System representation of ROM in the DQ domain.}
    \label{ROM}
\end{figure}

In our earlier works \cite{b10}, \cite{b16} and \cite{b17}, it was discussed that during grid faults, a type-4 WT could be reduced to a grid-side converter with a constant DC voltage (as seen in Fig. \ref{FullTop}a), where the control structure of the current controller and the phase-locked loop is presented in Fig. \ref{FullTop}b and Fig. \ref{FullTop}c, respectively. Furthermore, it was discussed that the fast inner current control dynamics could be neglected when analysing the slow PLL dynamics for large-signal stability analysis. Also, it was shown that the impact of the shunt capacitor filter could be neglected if the current is controlled on the grid side LCL filter. Considering all the assumptions, the grid-side converter can be represented by a controlled current source, wherein the current references are obtained from relevant grid codes. A reduced-order representation of the WT is illustrated in Fig. \ref{ROM}a, while the DQ-domain WT ROM under balanced grid conditions is presented in Fig. \ref{ROM}b.

From the DQ-domain WT ROM in Fig. \ref{ROM}b, the equivalent swing equation of the WT converter system derived in \cite{b10} can be presented as,

\begin{equation}\label{SEq_1}
M_{eq} \ddot{\delta} = T_{m_{eq}} - T_{e_{eq}} - D_{eq}\dot{\delta}
\end{equation}
where,
\begin{equation}\label{SEq_2}
\begin{aligned}
M_{eq} &= 1- k_p L_g i_d^c\\
T_{m_{eq}} 
&= k_p( \dot{\overline{r_{Lg} i_q^c}} + \ddot{\overline{L_g i_q^c}} + \dot{\overline{L_g i_d^c}} \omega_g)
+ k_i( r_{L_g} i_q^c \\
&\qquad+ \dot{\overline{L_g i_q^c}} + L_g i_d^c \omega_g) \\
T_{e_{eq}} 
&=  (k_i V_g \text{sin} \delta + k_p \dot{V_g} \text{sin} \delta) + M \dot{\omega}_g\\
D_{eq} 
&= k_p ( V_g \text{cos}\delta - \dot{\overline{L_g i_d^c}}) - k_i L_g i_d^c
\end{aligned}
\end{equation}

Equation (1) is a novel WT ROM that can represent the non-autonomous behaviour of the WT systems as validated in \cite{b10}. 
The ROM (1) was derived considering a balanced grid operation, i.e. to emulate grid faults, the voltage $v_f$ was varied. However, for unbalanced faults, the voltage $v_f$ must be computed considering the coupled sequence components. The methodology to extend the ROM (1) for unbalanced fault operation is discussed in Section II-C.

\subsection{Grid code requirements}
\begin{figure}[h]
    \centering
    \includegraphics[width=6.0cm]{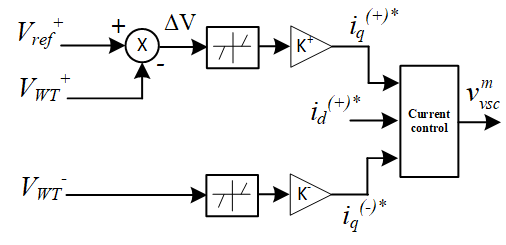}
    \caption{Typical grid code requirements - positive and negative sequence active and reactive current injection during grid faults.}
    \label{fig:RCI}
\end{figure}

As per grid code requirements \cite{b18}\cite{b19}, the WPPs are expected to stay connected even when the grid voltage is very low. This is achieved by the low voltage ride-through (LVRT) control systems, where in LVRT mode, new current references are generated based on grid code requirements. During unbalanced grid faults, both the positive and negative sequence currents should be controlled. The positive sequence reactive current injection is requested proportional to the positive sequence terminal voltage (see Fig. \ref{fig:RCI}). Similarly, negative sequence reactive current injection is requested proportional to the negative sequence terminal voltage. The positive and negative sequence gains can be adjustable between 2 and 6. The total reactive current is limited to 1pu. Thereafter, the active current is allowed to be regulated such that the total current is within the capability of the wind turbine.

\subsection{WT ROM based on symmetrical components}
As established from power system fault analysis, the positive, negative and zero sequences are interconnected at the fault through the fault impedance \cite{b20}. Due to this coupling, an action in one of the sequences also affects the other sequences, where the amount of boost in the couplings depends mainly on how much the WT can alter the voltage at the fault. 

\begin{figure}[h]
    \centering
    \includegraphics[width=5.75cm]{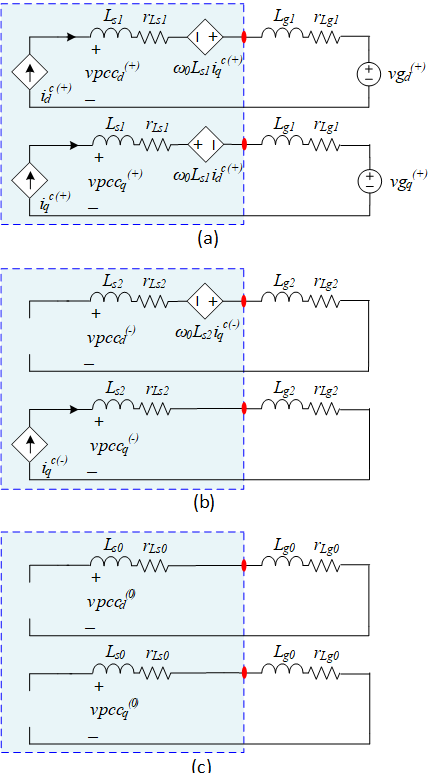}
    \caption{Sequence network for dq domain WT ROM: (a) positive sequence network, (b) negative sequence network, (c) zero sequence network.}
    \label{SeqNet}
\end{figure}

The pre-fault sequence network for DQ-domain WT ROM is presented in Fig. \ref{SeqNet}. It must be noted that here a grid impedance is introduced, which along with the fault impedance that will be later considered, is used to compute the unbalanced fault voltage. Further from grid codes, since there are no requirements of negative sequence active current injection or zero sequences current injection, the same is kept open circuit. 

\begin{figure}[h]
    \centering
    \includegraphics[width=7.8cm]{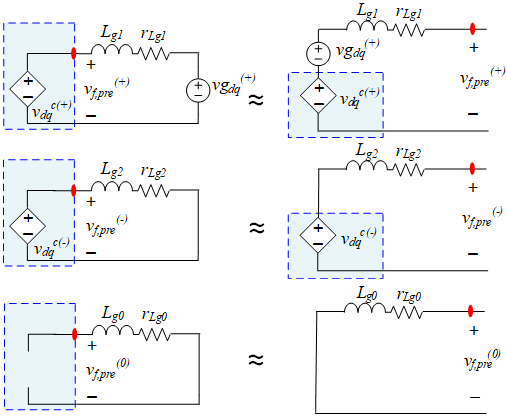}
    \caption{Simplified pre-fault sequence network diagram for WT ROM.}
    \label{SeqNet1}
\end{figure}

The sequence network for DQ-domain WT ROM can be further simplified as in Fig. \ref{SeqNet1}. Where, the pre-fault positive and negative sequence voltages $v_{f, pre}^{(+)}$ and $v_{f, pre}^{(-)}$ are given by (3) and (4), respectively.

\begin{figure*}[h]
    \centering
    \includegraphics[width=14.0cm]{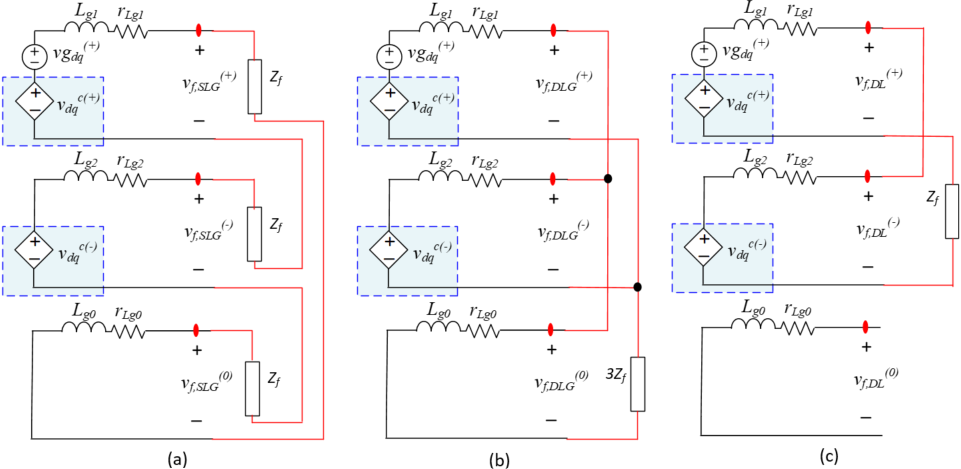}
    \caption{Post-fault sequence network diagram for WT ROM: (a) Single line to ground fault (ph-A), (b) Double line to ground fault (ph-B\&C), (c) Double line fault (ph-B\&C).}
    \label{summ}
\end{figure*}

\begin{equation}
\begin{aligned}
v_{f, pre}^{(+)} &= vg_{dq}^{(+)} + v_{dq}^{c(+)} \\
& = vg_{dq}^{(+)} + \left[ i_d^{(+)} + j\cdot i_q^{(+)}\right] \cdot \left[ r_{Lg1} + j\cdot \omega_0 L_{g1}\right]
\end{aligned}
\end{equation}
\begin{equation}
\begin{aligned}
v_{f, pre}^{(-)} &=  v_{dq}^{c(-)} \\
& = \left[ \cancelto{0}{i_d^{(-)}} + j\cdot i_q^{(-)}\right] \cdot \left[ r_{Lg2} + j\cdot \omega_0 L_{g2}\right]
\end{aligned}
\end{equation}

Once a grid fault is applied, the fault impedance couples all the sequence networks based on the type of grid fault. The coupled sequence network for different grid faults is presented in Fig. \ref{summ}. In the following section, the post-fault positive sequence voltages for various faults derived are presented.   

\subsubsection{Single line to ground fault}
The single line-to-ground (SLG) is the most common grid fault, which occurs when only one phase contacts the ground and has a non-zero fault current. Without loss of generality, we will assume it is phase A. The sequence diagram for the SLG fault is presented in Fig. \ref{summ}a, and the post-fault positive sequence voltage derived is presented in (5).
\begin{equation}\label{}
v_{f,SLG}^{(+)} = v_{f, pre}^{(+)} - \frac{v_{f, pre}^{(+)} + v_{f, pre}^{(-)}}{Z_{g1} + Z_{g2} + Z_{g0} + 3Z_f} \cdot Z_{g1}
\end{equation}
where, $Z_{g1} = (r_{Lg1} + j\cdot \omega_0 L_{g1})$, $Z_{g2} = (r_{Lg2} + j\cdot \omega_0 L_{g2})$ and $Z_{g0} = (r_{Lg0} + j\cdot \omega_0 L_{g0})$.

\subsubsection{Double line to ground fault}
Double line to ground occurs when two conductors come in contact with each other and the ground. Without loss of generality, we will assume phases b and c. The sequence diagram for the double line to ground fault is presented in Fig. \ref{summ}b, and the post-fault positive sequence voltage derived is presented in (6).

\begin{equation}\label{}
v_{f,DLG}^{(+)} = \frac{Z_{g0} + 3Z_f}{Z_{g2} + 2Z_{g0} + 6Z_f} \left[v_{f, pre}^{(+)} +v_{f, pre}^{(-)} \right]
\end{equation}

\subsubsection{Double line fault}
The second most common fault is a double line fault, which occurs when two conductors come in contact with each other. Similarly, without loss of generality, we will assume phases b and c.
The sequence diagram for the double line fault is presented in Fig. \ref{summ}c, and the post-fault positive sequence voltage derived is presented in (7).
\begin{equation}\label{}
v_{f,DL}^{(+)} = v_{f, pre}^{(+)} - \frac{v_{f, pre}^{(+)} - v_{f, pre}^{(-)}}{Z_{g1} + Z_{g2} + Z_{g0} + Z_f} \cdot Z_{g1}
\end{equation}

It must be noted that in existing literature, the not many works have analytically related the dependency of both positive and negative sequence DQ converter current injection on the fault voltages. Equations (3)-(7) addresses this issue, and is therefore a contribution of this paper.
Once the post-fault positive sequence voltages are computed via (3)-(7), the obtained voltages can be inputted to the WT ROM (1), and the resulting state trajectories can be obtained.

\subsection{Impact of Notch filter in PLLs}
A notch filter (NF) rejects frequencies inside a narrow band and allows other frequencies to pass without change. This property has made NFs a suitable tool for rejecting selective harmonic/disturbance components inside the PLL control loop. 

\begin{figure}[h]
    \centering
    \includegraphics[width=6.0cm]{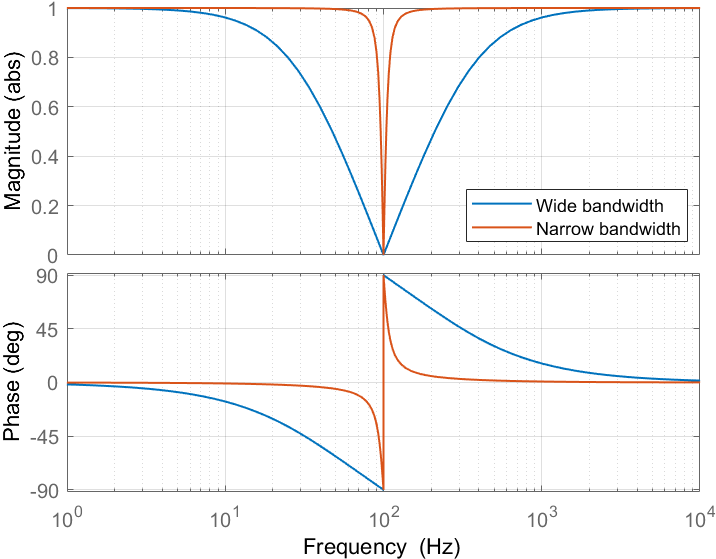}
    \caption{Bode plot for notch filters with wide and narrow bandwidth.}
    \label{NF}
\end{figure}

A major limitation with NFs in the PLL is that their rejection capability is significantly reduced with variations in the grid frequency. In such cases, the NFs dynamics must be considered which significantly complicates the PLL design. To address this challenge, using adaptive NFs is recommended \cite{b21} \cite{b22}. Frequency adaptability of NFs allows the PLL to reject the concerned disturbance components regardless of variations in the grid frequency. In addition, it enables a narrow bandwidth design for the NFs. It is known that a narrow bandwidth NFs causes a small phase delay in the PLL control loop and, therefore, their dynamics can be neglected without significantly affecting the PLL accuracy (seen in Fig. \ref{NF}).

In our proposed WT ROM, it is considered that notch filter is tuned to have a very narrow bandwidth that completely rejects the second harmonic oscillations (2$\omega_0$), as a result the dynamics of the notch filter can be safely neglected.

\section{Validation against PSCAD simulations}
In this section, the performance of the proposed ROM is analysed and is compared against an EMT simulation model of a WT in PSCAD \cite{b23}. Table 1 presents the WT system parameters considered in our study. In PSCAD model, the current controller is tuned to obtain a very fast response, while the PLL gains are chosen to attain an oscillatory behavior. The proposed ROM is tested against SLG, DLG and DL grid faults.

\begin{table}[h]
\caption{SYSTEM AND CONTROL PARAMETERS \cite{b23}}
\label{table}
\setlength{\tabcolsep}{3pt}
\begin{tabular}{p{40pt} p{125pt} p{50pt}}
\hline
Symbol& 
Description& 
Value \\
\hline
$S_b $& 
Rated power& 
12 MVA \\
$V_g$& 
Nominal grid voltage (L-N, pk) & 
690 $\sqrt{2/3}$ V\\
$V_{dc}$& 
DC-link voltage& 
1.38 kV \\
$f_0$& 
Rated frequency& 
50 Hz \\
$f_{sw}$& 
{Switching frequency} & 
10 kHz \\
$T_{s}$& 
{Simulation time step} & 
1 $\mu$s \\
$f_{s}$& 
{Sampling frequency} & 
1/$T_{s}$ \\
$r_c$, $L_c$& 
Converter-side inductor (pu) & 
0.005, 0.1 \\
$r_f$, $C_f$& 
Filter capacitor (pu) & 
0.0757, 0.3 \\
$r_{Ls}$, $L_{s}$& 
{Grid-side inductor (pu)} & 
0.0183, 0.3 \\
$r_{Lg}$, $L_{g}$& 
Grid impedance (pu) & 
0.0037, 0.06 \\
$i_d^c$, $i_q^c$& 
Pre-disturbance active and reactive currents (pu) & 
1.0, -0.1 \\
$K_{cc}$& 
Current controller gains & 
0.05, 0.3 \\
$K_{pll}$& 
SRF PLL gains & 
0.025, 1.5 \\
\hline
\end{tabular}
\label{tab1}
\end{table}

\subsection{Single line to ground fault}
\begin{figure}[h]
    \centering
    \includegraphics[width=7.50cm]{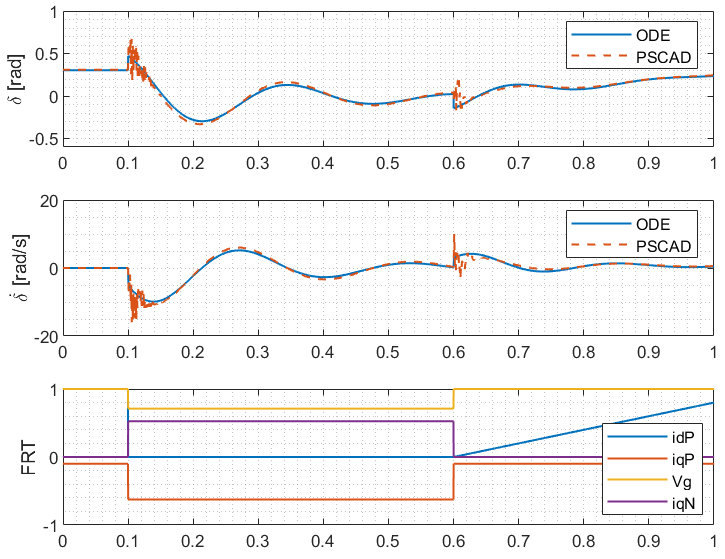}
    \caption{System trajectories with a SLG fault (ph-A) at the WT terminal.}
    \label{SLG}
\end{figure}

A SLG fault (ph-A) is applied at the WT terminal as shown in Fig. 1a. With a fault impedance of 6.02e-04 ohms, positive sequence active current injection of 0 pu, positive sequence reactive current injection of -0.625 pu, and negative sequence reactive current injection of 0.5 pu, the state trajectories of the PLL is presented in Fig. \ref{SLG}. 

It is seen that at the instant of disturbance, the PSCAD $\dot{\delta}$ has a slight overshoot and delay, this can be accounted for by the finite current controller bandwidth in PSCAD model, against the infinite current controller bandwidth considered in the WT ROM. However, the trajectories thereafter are similar to PSCAD simulations. In general a good match in $\delta$ and $\dot{\delta}$ trajectory is observed between the PSCAD simulations and proposed WT ROM.

\subsection{Double line to ground fault}
\begin{figure}[h]
    \centering
    \includegraphics[width=7.50cm]{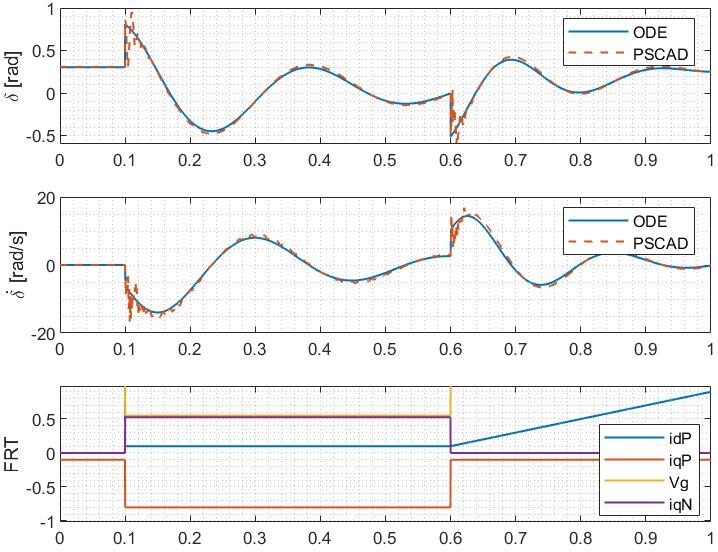}
    \caption{System trajectories with a DLG fault (ph-B\&C) at the WT terminal.}
    \label{DLG}
\end{figure}

Similarly, a DLG fault (ph-B\&C) is applied at the WT terminal as shown in Fig. 1a. With a fault impedance of 6.02e-04 ohms, positive sequence active current injection of 0.1 pu, positive sequence reactive current injection of -0.8 pu, and negative sequence reactive current injection of 0.2 pu, the state trajectories of the PLL is presented in Fig. \ref{DLG}. A good match in $\delta$ and $\dot{\delta}$ trajectory is observed between the PSCAD simulations and proposed WT ROM.

\subsection{Double line fault}
\begin{figure}[h]
    \centering
    \includegraphics[width=7.50cm]{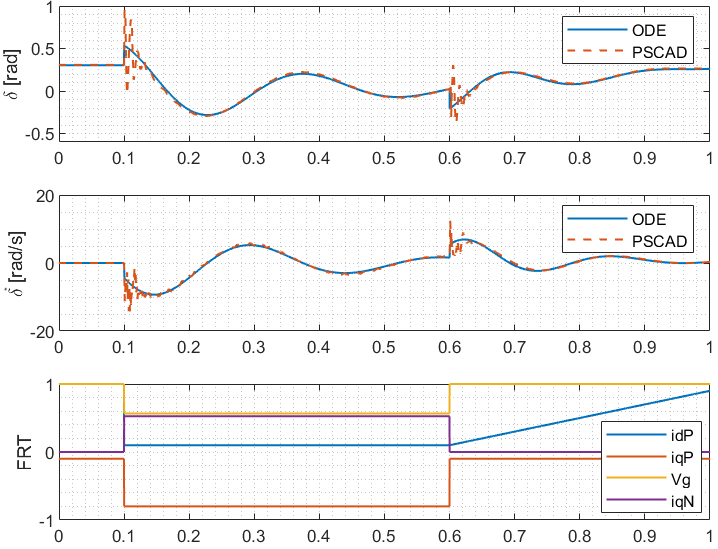}
    \caption{System trajectories with a  DL fault (ph-B\&C) at the WT terminal.}
    \label{DL}
\end{figure}

In the same way, a DL fault (ph-B\&C) is also applied at the WT terminal as shown in Fig. 1a. With a fault impedance of 6.02e-04 ohms, positive sequence active current injection of 0.1 pu, positive sequence reactive current injection of -0.8 pu, and negative sequence reactive current injection of 0.2 pu, the state trajectories of the PLL is presented in Fig. \ref{DL}. Overall, a good match in $\delta$ and $\dot{\delta}$ trajectory is observed between the PSCAD simulations and proposed WT ROM.

In general its is seen that the performance of the proposed ROM under unbalanced fault conditions is quite accurate over a wide range of frequencies owing to the consideration of non-autonomous behaviour. However, some deviations exist in the high-frequency range mainly driven by component modelling in PSCAD. For example, a slight delay with an overshoot in the PLL frequency is observed at the instant of a disturbance. This behaviour is introduced due to the current control loop (i.e. tuning a fast controller results in high overshoots, and the controller modelling inherently introduces a delay)\cite{b10}. Such high-frequency dynamics can be ignored when analysing the overall low-frequency synchronisation dynamics (i.e. the PLL acts as a low pass filter) of the PLL.

\section{Conclusion and future work}
A reduced-order model (ROM) for a type-4 wind turbine to investigate large signal stability under unbalanced grid faults was developed. The methodology presents a systematic way to model the coupled sequence components of the WT ROM for various grid faults. 
Based on the modelling and studies carried out in this paper, it is observed that post unbalanced grid disturbances, the proposed WT ROM correctly tracks the angle and frequency, and its trajectory is a good match when compared to a detailed PSCAD simulation model. Through time-domain simulations it was shown that neglecting the dynamics of the notch filter in the PLL is a valid consideration as long as the filter is designed to have a narrow bandwidth.  

Some possible future work of the WT ROM is to estimate the frequency range of the proposed model, and consider impact of different filter topology, to filter out the 2nd order harmonics, during grid faults.   






\vspace{12pt}


\begin{thebibliography}{00}
\bibitem{b1} IEA (2022), Wind Electricity, IEA, Paris https://www.iea.org/reports/wind-electricity, License: CC BY 4.0.1955.

\bibitem{b2} Y. Huang, X. Yuan, J. Hu and P. Zhou, "Modeling of VSC Connected to Weak Grid for Stability Analysis of DC-Link Voltage Control," in IEEE Journal of Emerging and Selected Topics in Power Electronics, vol. 3, no. 4, pp. 1193-1204, Dec. 2015.

\bibitem{b3} C. Zhang, X. Cai, A. Rygg, and M. Molinas, “Sequence domain SISO equivalent models of a grid-tied voltage source converter system for smallsignal stability analysis,” in IEEE Trans. Energy Convers., vol. 33, no. 2, pp. 741–749, Jun. 2018.


\bibitem{b5} M. Bravo, A. Garcés, O. D.Montoya, and C. R. Baier, “Nonlinear analysis for the three-phase PLL: A new look for a classical problem,” in Proc. IEEE Workshop Control Model. Power Electron., Padua, Italy, Jun. 2018, pp. 1–6.

\bibitem{b6} L. Harnefors, X. Wang, A. Yepes, and F. Blaabjerg, ``Passivity-basedstability assessment of grid-connected VSCs-an overview,'' \emph{IEEE J. Emerg. Select. Topics Power Electron}, vol. 4, no. 1, pp. 116–125.

\bibitem{b7} Y. Gu, N. Bottrell, and T. C. Green, ``Reduced-order models for representing converters in power system studies,'' \emph{IEEE Trans. Power Electron}, vol. 33, no. 4, pp. 3644–3654, Apr. 2018.

\bibitem{b8} D. Dong, B. Wen, D. Boroyevich, P. Mattavelli and Y. Xue, ``Analysis of Phase-Locked Loop Low-Frequency Stability in Three-Phase Grid-Connected Power Converters Considering Impedance Interactions,'' \emph{IEEE Transactions on Industrial Electronics}, vol. 62, no. 1, pp. 310-321.

\bibitem{b9} H. Geng, L. Liu, and R. Li, ``Synchronisation and reactive current support of PMSG-based wind farm during severe grid fault,'' IEEE Trans. Sustain. Energy, vol. 9, no. 4, pp. 1596–1604, Oct. 2018.

\bibitem{b10} Mohammad Kazem Bakhshizadeh, Sujay Ghosh, Łukasz Kocewiak, and Guangya Yang. (2022). Improved Reduced-Order Model for PLL Instability Investigations. https://doi.org/10.5281/zenodo.7016303

\bibitem{b11} V.D. Bacon, S.A. Oliveira da Silva, Performance improvement of a three-phasephase-locked-loop algorithm under utility voltage disturbances usingnon-autonomous adaptive filters, IET Power Electron. 8 (November (11))(2015) 2237–2250.

\bibitem{b12} A. Kulkarni, V. John, Design of synchronous reference frame phase-lockedloop with the presence of dc offsets in the input voltage, IET Power Electron. 8(December (12)) (2015) 2435–2443.

\bibitem{b13} M.E. Meral, Improved phase-locked loop for robust and fast tracking of threephases under unbalanced electric grid conditions, IET Gene. Transm. Distrib. 6(February (2)) (2012) 152–160.

\bibitem{b14} X. Guo, W. Wu, Z. Chen, Multiple-complex coefficient-filter-basedphase-locked loop and synchronization technique for three-phase gridinterfaced converters in distributed utility networks, IEEE Trans. Ind. Electron.58 (April (4)) (2011) 1194–1204.

\bibitem{b15} F.D. Freijedo, J. Doval-Gandoy, O. Lopez, E. Acha, Tuning of phase locked loopsfor power converters under distorted utility conditions, IEEE Trans. Ind. Appl.45 (November/December (6)) (2009) 2039–2047.

\bibitem{b16} S. Ghosh,  M. K. B. Dowlatabadi, G. Yang, and L. Kocewiak, "Non-linear Stability Boundary Assessment of Offshore Wind Power Plants Under Large Grid Disturbances," 21st International Workshop on Large-Scale Integration of Wind Power into Power Systems as well as on Transmission Networks for Offshore Wind Power Plants (WIW 2022).

\bibitem{b17}
S. Ghosh, G. Yang, M. K. B. Dowlatabadi and L. Kocewiak, "Nonlinear stability analysis of a reduced-order wind turbine VSC-grid model operating in weak grid conditions," 20th International Workshop on Large-Scale Integration of Wind Power into Power Systems as well as on Transmission Networks for Offshore Wind Power Plants (WIW 2021).

\bibitem{b18} "Technical regulation 3.2.5  for wind power plants above 11 kW", Energinet 13/96336-43.

\bibitem{b19} “Offshore-Netzanschlussregeln (O-NAR)”, Tech. rep. TenneT TSO GmbH, August 2019. 

\bibitem{b20} J. C. Das, “Power system analysis: short-circuit load flow and harmonics”, Marcel
Dekker Press, 2002.

\bibitem{b21} F. Gonzalez-Espin, G. Garcera, I. Patrao, E. Figueres, An adaptive controlsystem for three-phase photovoltaic inverters working in a polluted andvariable frequency electric grid, IEEE Trans. Power Electron. 27 (October (10))(2012) 4248–4261.

\bibitem{b22} S. Eren, M. Karimi-Ghartemani, A. Bakhshai, Enhancing the three-phasesynchronous reference frame PLL to remove unbalance and harmonic errors,in: Proc. 35th Annu. Conf. IEEE Ind. Electron., November, 2009, pp. 437–441.

\bibitem{b23} Ł.  Kocewiak,  R.  Blasco-Giménez,  C.  Buchhagen,  J.  B.  Kwon,  Y.  Sun, A.  Schwanka  Trevisan,  M.  Larsson,  X.  Wang,  “Overview,  Status  and Outline  of  Stability  Analysis  in  Converter-based  Power  Systems,”  The 19th International Workshop on Large-Scale Integration of Wind Power into Power Systems. as well as Transmission Networks for Offshore Wind Farms, Energynautics GmbH, 11-12 November 2020.

\end{thebibliography}
\end{document}